\documentclass[useAMS,usenatbib, letters]{mnras}

\usepackage{graphicx}   % Including figure files
\usepackage[english]{babel}

\newcommand{\OIII}{[O~{\sc iii}]\ }
\newcommand{\NII}{[N~{\sc ii}]\ }

\newcommand{\Ha}{H$\alpha$\ }
\newcommand{\Hb}{H$\beta$\ }
\newcommand{\kms}{\,\mbox{km}\,\mbox{s}^{-1}}
\newcommand{\ergs}{\,\mbox{erg}\,\mbox{s}^{-1}}

\title[Emission filaments in Mrk~6]{A close look at the well-known Seyfert galaxy: extended emission filaments in Mrk~6
\thanks{Based on observations obtained with the 6-m telescope of the Special Astrophysical Observatory of the Russian Academy of Sciences carried out with the financial support of the Ministry of Education and Science of the Russian Federation (agreement No. 14.619.21.0004, project ID RFMEFI61914X0004). }
}

\author[Smirnova, Moiseev \& Dodonov]
{Aleksandrina A. Smirnova $^{1}$\thanks{E-mail:ssmirnova@gmail.com} , 
Alexei V. Moiseev $^{1,2}$\thanks{E-mail:moisav@gmail.com} ,
 and Sergei N. Dodonov $^{1}$ \\
 $^{1}$ Special Astrophysical Observatory, Russian Academy of Sciences,  369167 Nizhny Arkhyz,  Karachaevo-Cherkesskaya Republic,  Russia\\
 $^{2}$ Space Research Institute, Russian Academy of Sciences, Profsoyuznaya ul. 84/32, Moscow 117997, Russia
 }

\date{Accepted 2018 Month 00. Received 2018  Month 00; in original
form 2018 Month 00}

\pubyear{2016}

\begin{document}
 
\label{firstpage}
\pagerange{\pageref{firstpage}--\pageref{lastpage}}
\maketitle

\begin{abstract}
Using various techniques of optical observations at the 6-m Russian telescope (emission-line images, long-slit and 3D spectroscopy), we have studied large-scale morphology and kinematics of the ionized gas in the Seyfert galaxy Mrk~6. Having  a significantly deeper images and spectra than those in previous works, we have not only mapped the ionized gas in the stellar disc but also found a system of faint filaments elongated  in the NE direction up to a projected distance of 40 kpc ($4.3R_{25}$).   Kinematics as well as an ionization state of the filament gas suggest the scenario that  hard radiation of the active nucleus illuminated the matter accreted from outside and orbiting  almost orthogonally to the Mrk~6 stellar disc. A possible source of gas accretion is still unknown, a deep image taken with the 1-m Byrakan Schmidt telescope does not show any stellar counterparts  at the level of 27--28 mag~arcsec$^{-2}$.

\end{abstract}


\section{Introduction}

The central engines of active galactic nuclei (AGN) have a significant  impact on the ambient interstellar medium via relativistic jets, gas outflow, and collimated hard ionizing radiation. In some cases,   Extended Emission-Line Regions (EELRs) are observed at as far as tens kiloparsecs. In radio-quiet  objects including local Seyfert galaxies, the jet-cloud interaction occurs only in the central several kiloparsecs, while the rest parts of an EELR ionized by the AGN radiation appear  in a broad cone. Recent works used EELRs to probe the history of an AGN radiative output across the light-travel times to the external gaseous clouds  \citep[`a quasar light echo',][]{Lintott2009,Keel2012} that is also called the `AGN archeology' \citep{Morganti2017}. 
The other aspect of the EELR study related with the AGN illumination of gas outside the main galaxy disc: off-plane tidal debris \citep{Keel2015} or even  cosmic web filaments  \citep{Cantalupo2014}. Therefore, the study of EELRs highlights the processes of baryon mass assembly by galaxies with significantly higher angular resolution than that in radio observations of neutral H~{\sc i}. In this Letter, we described the system of extended filaments around the Mrk~6 AGN  serendipitously discovered in the scanning Fabry-Perot Interferometer (FPI) observations at the 6-m telescope of the Special Astrophysical Observatory of the Russian Academy of Sciences (SAO RAS).

\begin{table*}
\caption{Log of observations at the SAO RAS 6-m telescope}
\label{tab_log}
\begin{tabular}{@{}llllll@{}}
\hline
Date      & Instrument/mode & Exp. time, s              & Sp. range, \AA     &   Sp.  resol. (FWHM), \AA            &       Seeing, arcsec         \\
%                                &                                                   &      (s)                    &  \AA              &   (FWHM), \AA   &   arcsec           \\
\hline
%14 Feb 2007   &  SCORPIO/IM                             &7$\times$600  & H$\alpha_{on}$&                 &   1.6     \\
 %                       &                                                     & 1200      & H$\alpha_{off}$ &                 &   1.6      \\
22/23 Feb 2006&SCORPIO/FPI  &32$\times$300  & H$\alpha$      &       2.6     &  1.8--2.2    \\
16/17 Dec 2014&SCORPIO-2/IM &7$\times$600    & H$\alpha_{on}$&               &   1.6     \\
              &             & 7$\times$180  & H$\alpha_{off}$&               &   1.6      \\
05/06 Mar 2014&SCORPIO-2/IM &5$\times$300  &[OIII]$_{on}$    &               &2.6         \\
              &             &5$\times$300   &[OIII]$_{off}$  &               &2.6           \\
01/02 Apr 2016&SCORPIO-2/IM & 5$\times$120  & $g$-sdss       &               &   1.8     \\
              &             & 4$\times$90   & $r$-sdss       &               &   1.6      \\
              &             & 5$\times$150  & $i$-sdss       &               &   1.7      \\                        
05/06 Mar 2014& SCORPIO-2/LS PA=144$\degr$&2$\times$900& 3650--7230&    4.5            & 2.4              \\
              & SCORPIO-2/LS PA=226$\degr$&2$\times$900 & 3650--7230&    4.5            & 2.8              \\
02/03 Dec 2000& MPFS         &3$\times$1200 & 3750--6370     &    8.0        &2.0         \\
                                                        
\hline
\end{tabular}
\end{table*}

Mrk~6 is a well-known S0  galaxy identified as an intermediate Seyfert through the optical spectroscopy by \citet{Osterbrock}.
%Although it is extensively studied spectroscopically, but deep direct images information about this object is very poor. 
It has a strongly variable continuum and broad Balmer lines 
\citep[and references therein]{Khachikian,Eracleous,Doroshenko2012} extensively studied spectroscopically, the mass of the central black hole is about  $1.5\times10^8$ M$_{\odot}$ according to the spectropolarimetry by \citet{Afanasiev2014}. Mrk~6 reveals a complex structure of radio emission: a compact radio core and elongated multicomponent radio structures which resemble jet elements and/or hot spots. It is also one of Seyfert galaxies that have radio emission on both small and large ($\sim15$ kpc in diameter) scales \citep{Kukula1996,Kharb2014}, its complex radio emission is argued to have been possibly originated by an episodically powered precessing jet  \citep{Kharb2006}.  HST images also show jet-like emission feature  extending at $\sim200$ pc, which is co-spatial with a radio jet seen in the 6-cm radio data by \citet{Capetti1995}. The ground-based imaging and spectral observations revealed a large ($r\approx22$ kpc) EELR  { with  a multi-component structure of the  emission-line profiles  in the circumnuclear ($r<3$ kpc) region,  whereas the gas emission outside demonstrates a single kinematic component with a low velocity dispersion   \citep*{Meaburn1989}. \citet{AfanaSil1991} used optical spectroscopic data taken in five  different long-slit positions together  with integral-field observations to reconstruct the ionized gas kinematics of EELR,  the observed picture was very complex: the authors decided that the inner region has a possible opposite rotation with significant outflow motions, whereas the other part of the gaseous disc is inclined by $40\degr$ to the galaxy's line-of-sight.  These controversial kinematics} was the reason to select this  target for our 3D spectroscopic observations. We   adopted the distance to the  galaxy  as 80.6 Mpc \citep{Kharb2006} with the scale of 0.39 kpc per arcsec. 

% This non-interacting galaxy has distorted outer parts. 
%The presence of dust lanes northern to the nucleus, stellar shells at the eastern and western edges of the galaxy, knots and filaments on H$\alpha$ images and surrounding extended emission were registered by \citet{Khachikian,Munoz}. %\citet{Cracco}. 

\begin{figure*}
\centerline{
\includegraphics[width=6 cm]{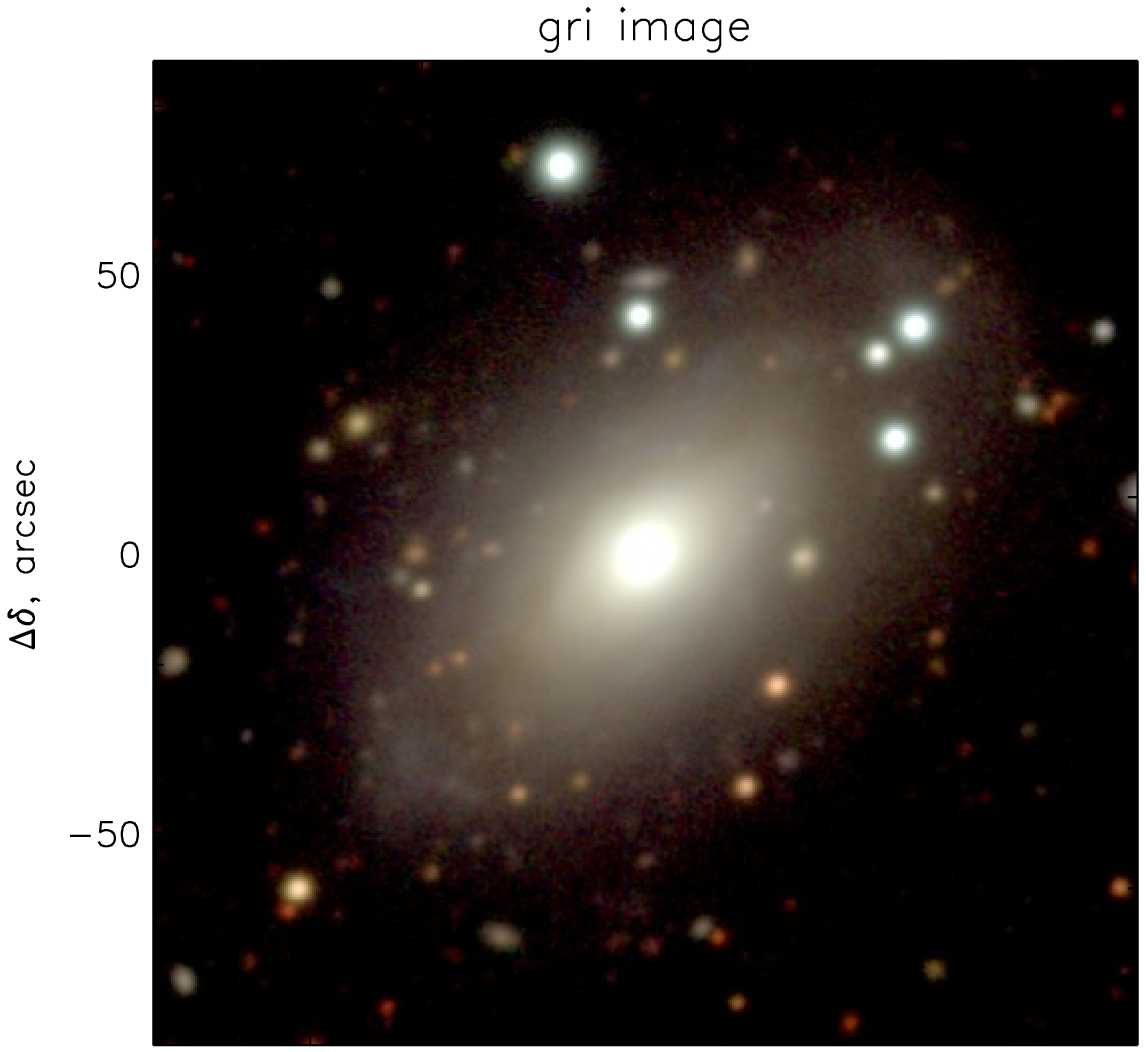}
\includegraphics[height=5.6cm]{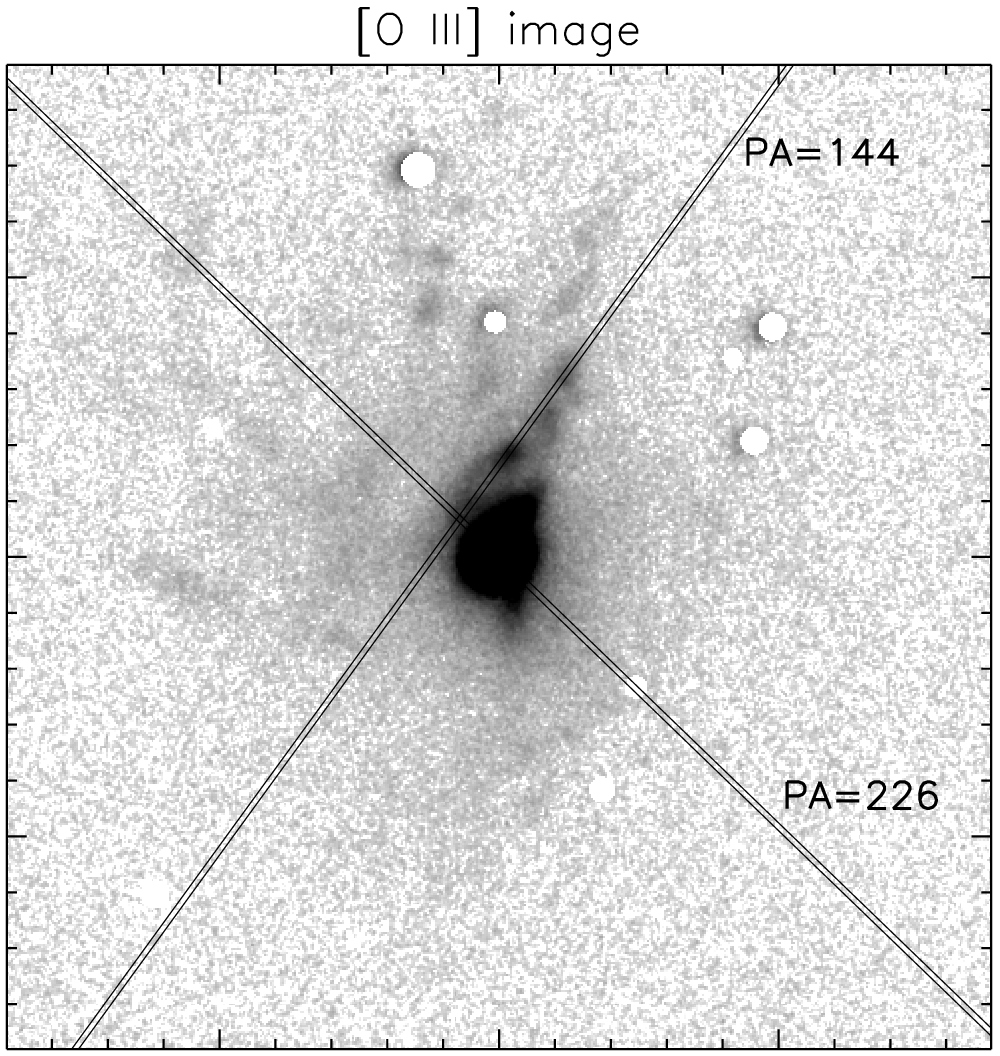}
\includegraphics[height=5.6 cm]{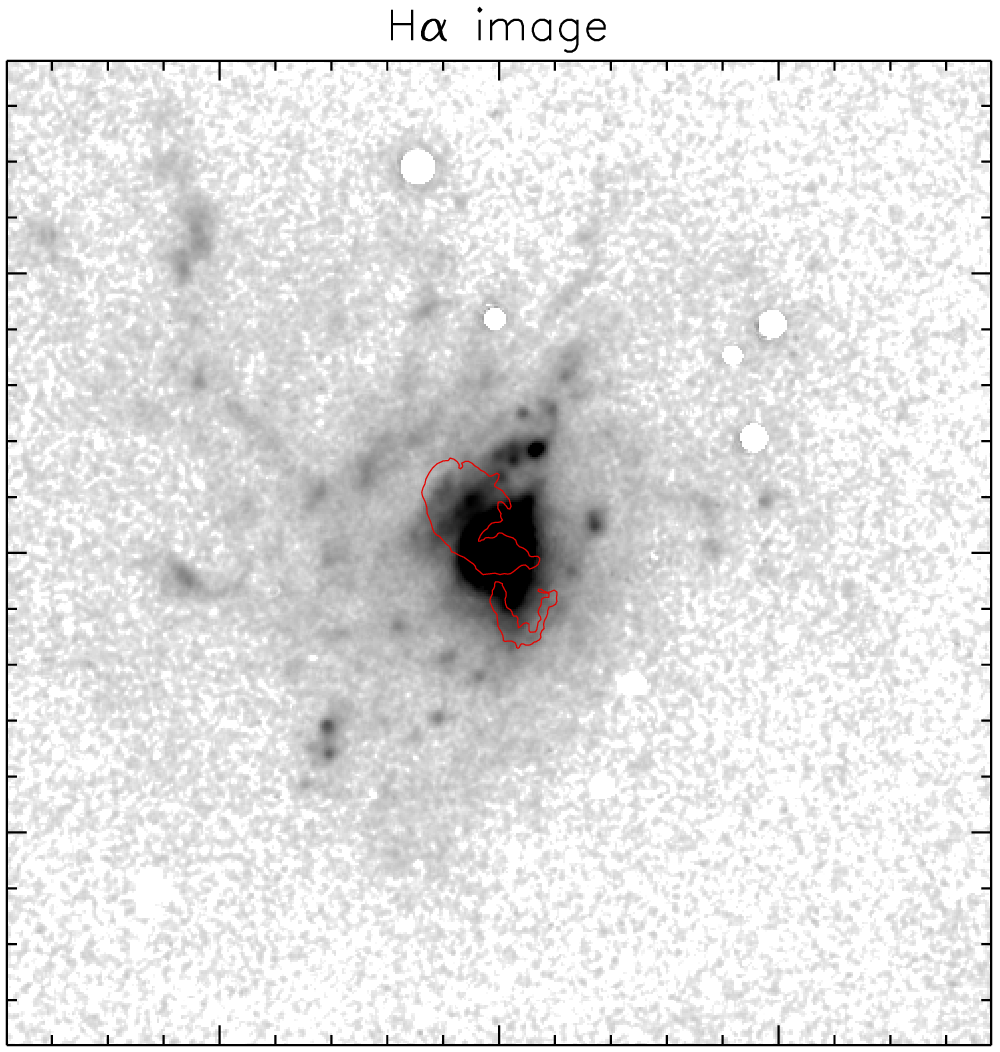}
}
\centerline{
\includegraphics[width=6 cm]{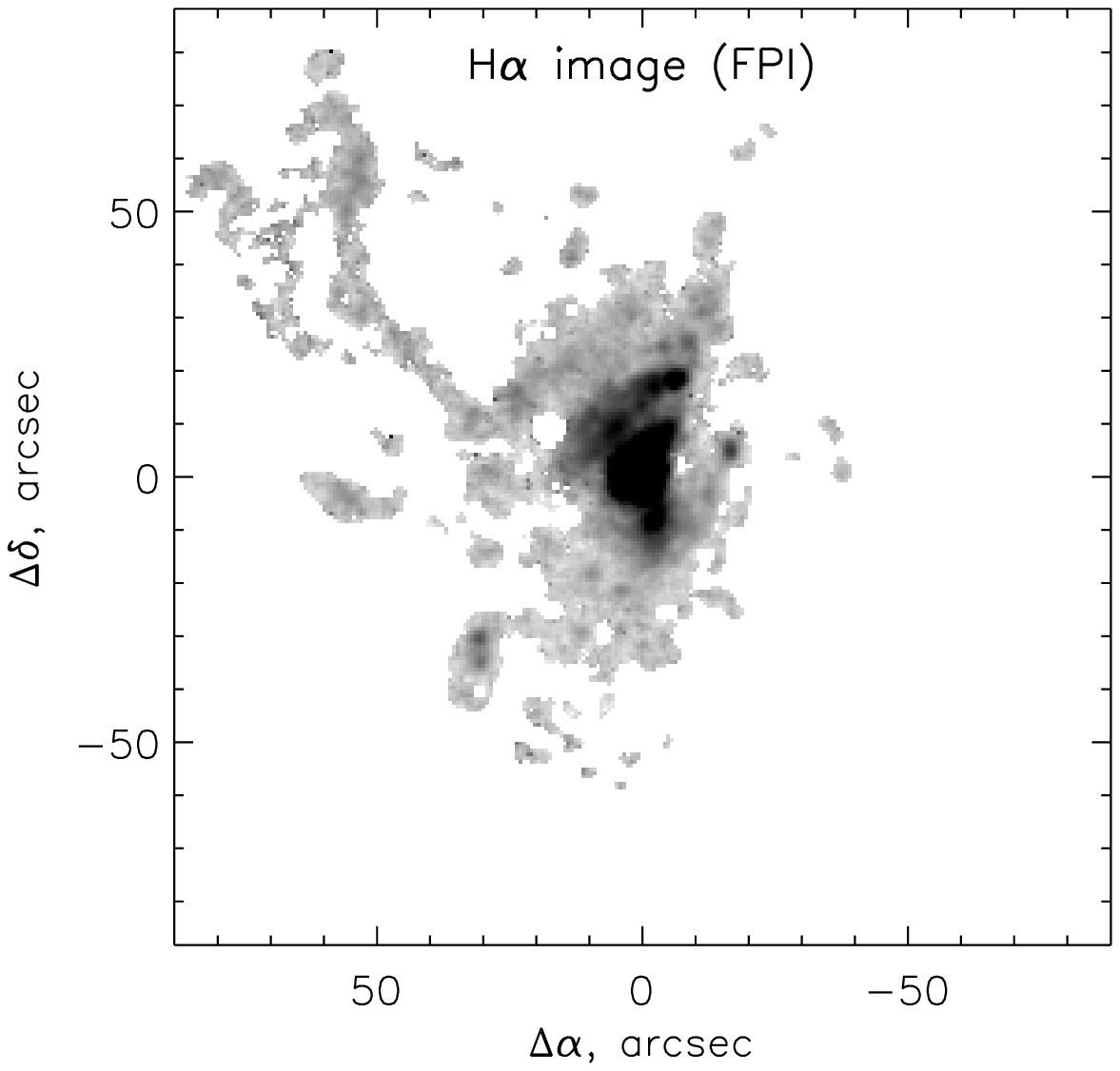}
\includegraphics[width=5.1  cm]{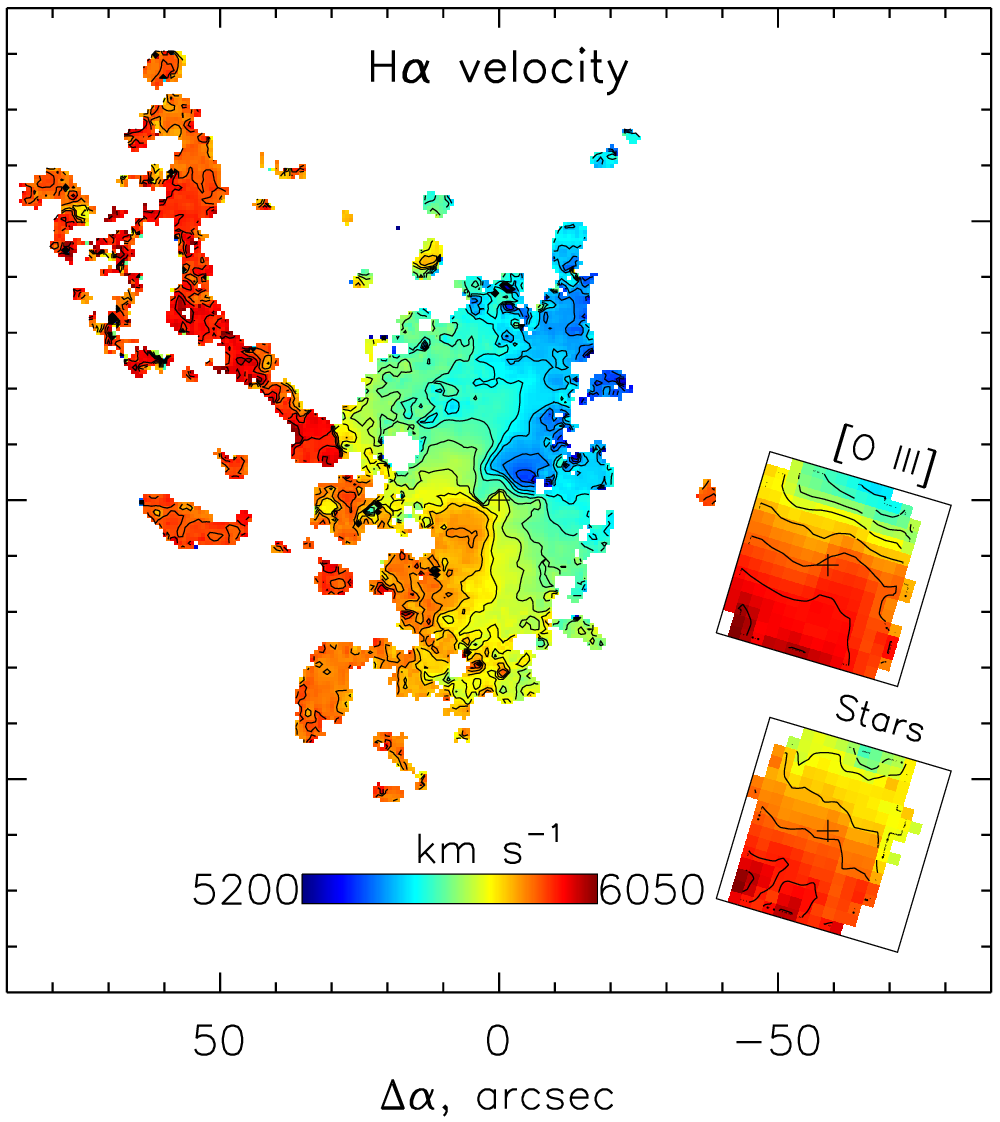}
\includegraphics[width=5.1 cm]{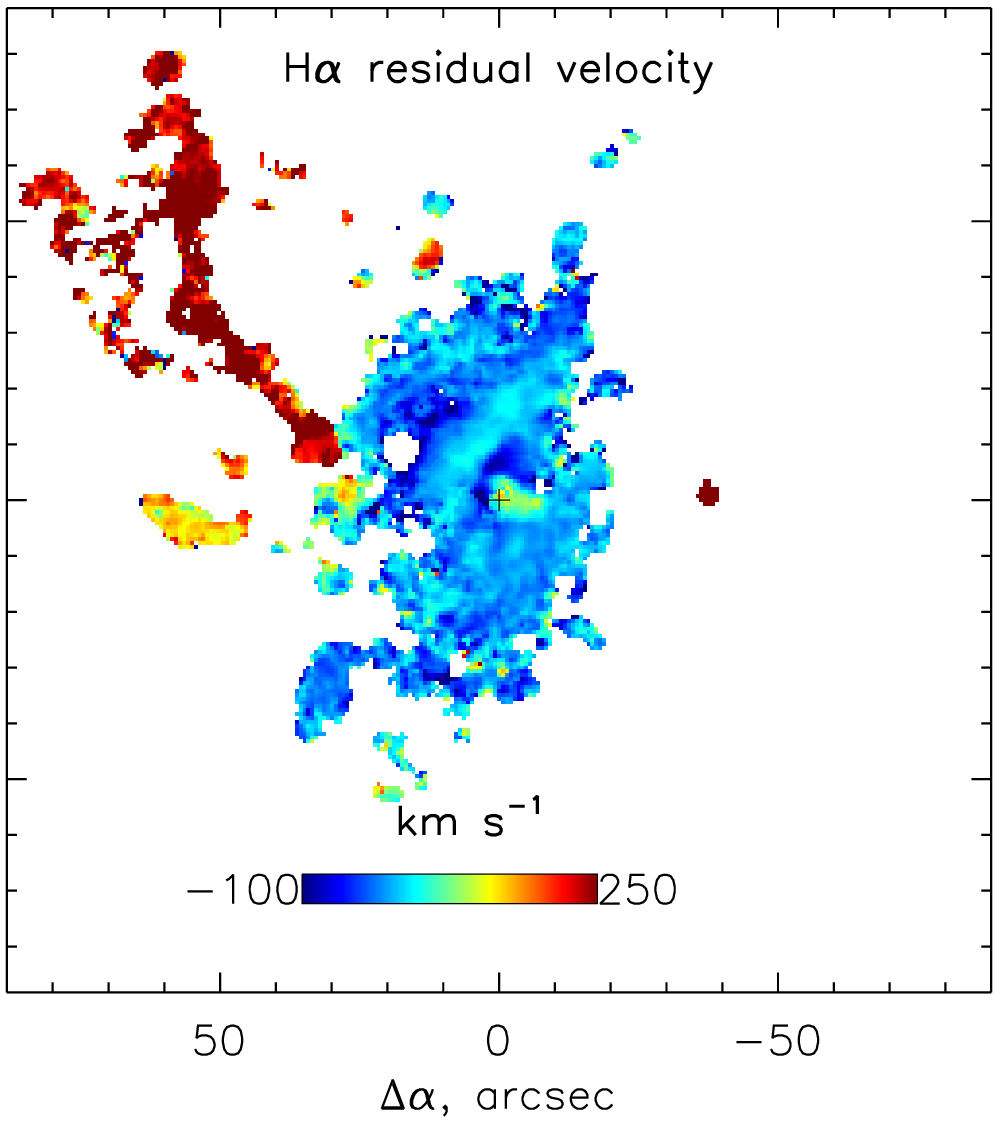}
}
\caption{Mrk~6 data obtained at the 6-m telescope. Direct images are in the top row: the {\it gri} composite image (left-hand); the [OIII] emission line image with marked slit positions (middle); the \Ha  image (right-hand), the red contours mark the external border of the radio structure   on 20 cm \citep{Kharb2006}. 3D spectroscopic data are in the bottom row: the map of the \Ha emission obtained from the  lines  fitting the FPI data cube (left-hand); the line-of-sight velocities in \Ha according to FPI observations together with the circumnuclear ($16\times16$ arcsec$^2$) MPFS velocity fields of stars and the ionized gas  in \OIII (middle); the residual  \Ha velocities after subtraction of the disc circular rotation model (right-hand). 
%Scale boxes are in $\kms$. 
}
\label{fig_maps}
\end{figure*}

\section{Observations and data reduction}
\label{sec_obs}
Observations were carried out at the prime focus of the 6-m SAO RAS telescope in different modes of the  SCORPIO \citep{AfanasievMoiseev2005} and SCORPIO-2 \citep{SCORPIO2} focal reducers in 2006--2016. Also, we used the 6-m telescope archival data  obtained with the integral-field Multi-Pupil Field Spectrograph \citep*[MPFS,][]{Afanasiev2001}. The log of observations and parameters of the used instruments are listed in  Table~\ref{tab_log}. The main steps of  data reduction and references to the  software are presented in our previous papers \citep{Smirnova2010,Egorov2018}.  SCORPIO/SCOPRIO-2 provided 6.1-arcmin  field of view (and the same length of the long slit with the $1$-arcsec width for  spectroscopy) with the 0.35-arcsec pixel scale. The scanning FPI mapped the spectral region around the redshifted \Ha emission line in 32 spectral channels. We made additional binning  of these data to reach a higher signal-to-noise (S/N) ratio in the resulting   data cube with $0.7$ arcsec per pixel.  The MPFS data cube covered the larger spectral range in the $16\times16$ arcsec$^2$ field-of-view centered at the nucleus, the spatial scale was 1 arcsec. 

The narrow-band  images were taken through filters centered on the redshifted \Ha and [O~{\sc iii}]$\lambda5007$ emission lines and the nearest continuum. The corresponding filters are marked with the indexes `on' and `off' in   Table~\ref{tab_log}. The  FWHM  filters  were  79\AA\ for the H$\alpha_{off}$ filter and 15--33\AA\ in  other cases. Calibration of these narrow-band images as well as the SCORPIO-2 and MPFS spectra  into the absolute energetic flux units was performed using spectrophotometric standard stars observed in the same nights as Mrk~6. 
In order to convert the broad-band images in the SDSS $gri$ filters into the  scale of magnitudes, we used the photometric catalogue of the   Pan-STARRS1  field stars \citep{Magnier2016}. 

Also, deep images of the $\sim1\degr$  field centered at Mrk~6 were obtained at the 1-m Schmidt telescope (Byurakan Astrophysical Observatory (BAO) of the National Academy of Sciences of Armenia) in the $gri$ SDSS filters in Nov 5--6, 2016. The total exposure was 55 min  in  each filter with a typical  seeing of $\sim3$ arcsec with the 0.87-arcsec pixel scale. The description of this camera and the data reduction process were presented by \citet{Dodonov2017}. Calibration of the images into  magnitudes was performed in the same manner as for the SCORPIO-2 broad-band images.

\begin{figure}
\centerline{
\includegraphics[width=0.45\textwidth ]{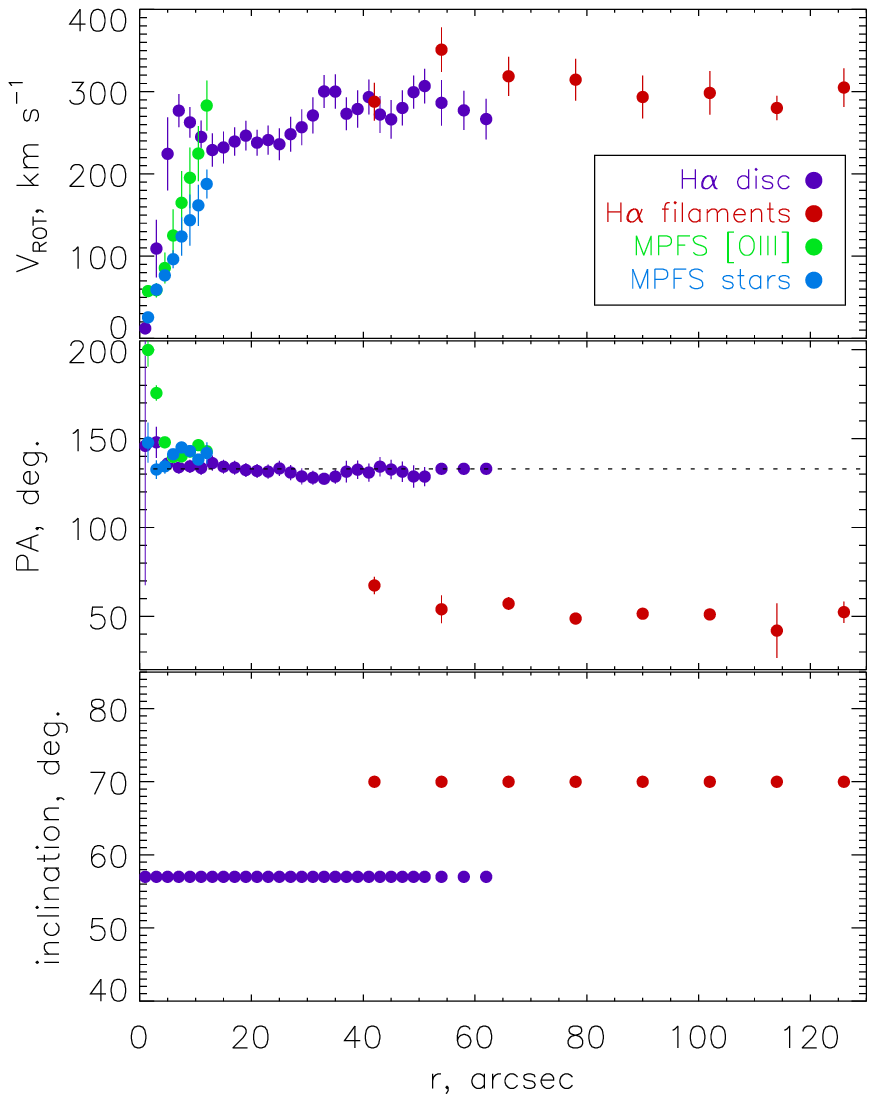}
}
\caption{Radial variations of the tilted-ring model  parameters: the circular rotation
velocity (top),  the   position angle {(middle) and the accepted inclination (bottom).} The dashed line marks the $PA$ of the galaxy's stellar disc.}
\label{fig_rc}
\end{figure}

\section{Morphology and gas kinematics}
\label{sec_morph}
Figure~\ref{fig_maps} shows the composite $gri$ SCORPIO-2 image of the galaxy  that is significantly  deeper than the broad-band images published previously. The $1\sigma$ noise  level in background pixels corresponds to  surface brightnesses of 27.7, 27.4, and 27.2 mag~arcsec$^{-2}$ in the $g,r,$ and $i$ filters, respectively. This image does not reveal any considerable peculiarities (tidal features or asymmetric structures)  in  the distribution of the stellar component. The only exception is low-contrast arches %(\red{$\Delta m\approx0.5^m$  ??}) 
or a pseudo ring at the distances $r\approx60$ arcsec near the galaxy's major axis. However, the observed morphology  changes dramatically in the ionized-gas emission lines. The 6-m telescope narrow-band images and FPI maps clearly demonstrate that the EELR known early and extending at $r=13$--22 kpc  \citep{Meaburn1989,Kukula1996} is only an inner part of a giant system of  knots  and diffuse filaments. The filaments were detected up to a surface brightness level of $\sim5\cdot10^{-18}\ergs$arcsec$^2$ in the \Ha emission line and  $\sim1\cdot10^{-17}\ergs$arcsec$^2$ in  \OIII that is six times fainter than those in the \citet{Meaburn1989} images. The most extended filaments elongate in the NE direction  at the projected distance $r=102$ arcsec  ($\sim40$ kpc) that corresponds to $4.3R_{25}$ ($R_{25}=24$ arcsec according to the NED database).

The bottom panels in Fig.~\ref{fig_maps} show the maps derived from the one-component Voigt fitting of the \Ha spectra in the FPI data cube. The distribution of emission structures in the FPI \Ha monochromatic map agrees very well with the narrow-band direct image  if some difference in the detection limit are taken into account.  

The inner part of the \Ha velocity field  ($r<60$ arcsec) has a pattern typical of dominating circular rotation. To describe the gas kinematics quantitatively,  we fitted the velocity field with a classical `tilted-ring' model using the technique described in  \citet[][and references therein]{Smirnova2010,Moiseev2014}. We kept the disc inclination along the radius as $i=57\degr$ according to our estimation of the outer $g$- and $r$-band isophotes. 

The radial changes of the rotation velocity $V_{rot}$ and major kinematic axis $PA_{kin}$ are shown in Fig.~\ref{fig_rc}. The   \Ha velocity field might be divided into two components: (i) the galaxy disc and (ii) the external filaments. In the inner gaseous disc, the radial variation of $PA_{kin}$ around its mean value of $133\pm3\degr$ are negligible up to $r\approx60$ arcsec (22 kpc)  that agrees very well with the position angle of the stellar continuum isophotes. The  velocity field of the stellar component  in the circumnuclear ($r<10$ arcsec) region derived from the MPFS data cube using the cross-correlation technique \citep[as described in][]{Smirnova2010} also has the same  $PA_{kin}$. Therefore, the gaseous disc co-rotates with the stellar one. The typical value  of  residual  velocities after subtraction  of the tilted-ring model is about $\pm30\kms$, while they reach an amplitude of 85--120$\kms$ in several regions near the nucleus at $r<9$ arcsec. These non-circular motions seem to be related with the AGN outflow and   jet-cloud interaction and located near the edges of the radio structure as presented in \citet{Kharb2006}.  Additional evidence  of the AGN influence on the surrounding-gas kinematics is provided by significant variations of  $PA_{kin}$ estimated in the high-excited \OIII emission lines in our MPFS data (the MPFS maps are shown in Fig.~\ref{fig_maps} as insets in the panel with the \Ha velocity field). Our spatial resolution is insufficient for a detailed study of gas properties in these regions  which were recently mapped  with the sub-arcsecond integral-field observations by \citet{Freitas2018}. The authors also stressed the distinct rotation pattern  in the \OIII and \Ha velocity fields related to gas compression by the jet. 
The multicomponent structure of  narrow lines in this region was also mentioned by \citet{Meaburn1989} and \citet{AfanaSil1991}.

The kinematics of the external gaseous filaments NE to the nucleus is fully decoupled from the disc circular rotation. Indeed, the map of residuals in Fig.~\ref{fig_maps} shows that the observed line-of-sight velocities exceed the extrapolation of the circular rotation model by 100--300$\kms$. It is interesting that this peculiar kinematics also can be described in a `tilted-ring' model of rotation, if we assume the same systemic velocity as that for the galaxy's disc and  inclinations of about $70\degr$. The model solution for this case seems to be very reasonable (the red symbols in Fig.~\ref{fig_rc}): a flat rotation curve with the same amplitude as that for the Mrk~6 disc up to $r=125$ arcsec and a relatively small variation  $PA_{kin}\approx50\degr$ in the agreement with the observed extension of the filaments. In this case, the external gas rotates around Mrk~6 in the polar plane almost orthogonally to the  galaxy's stellar disc.

\section{Gas excitation}

\begin{figure}
\centerline{
\includegraphics[width=0.5\textwidth ]{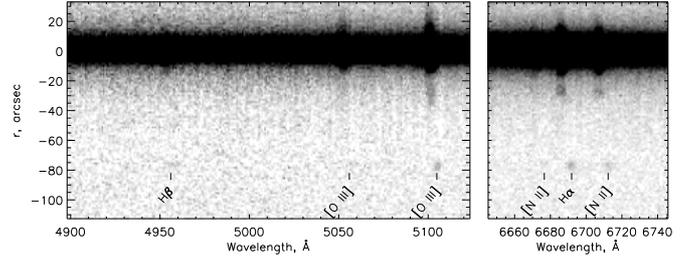}
}
\caption{Mrk~6 spectrum along $PA=226\degr$:   fragments of the 2D spectrum around the H$\beta$+[OIII]$\lambda\lambda4959,5007$ and \Ha+[NII]$\lambda\lambda6548,6583$  regions binned to a   scale of 1.4 arcsec  along the slit.}
\label{fig_spec}
\end{figure}

\begin{figure}
\centerline{
\includegraphics[width=0.45\textwidth ]{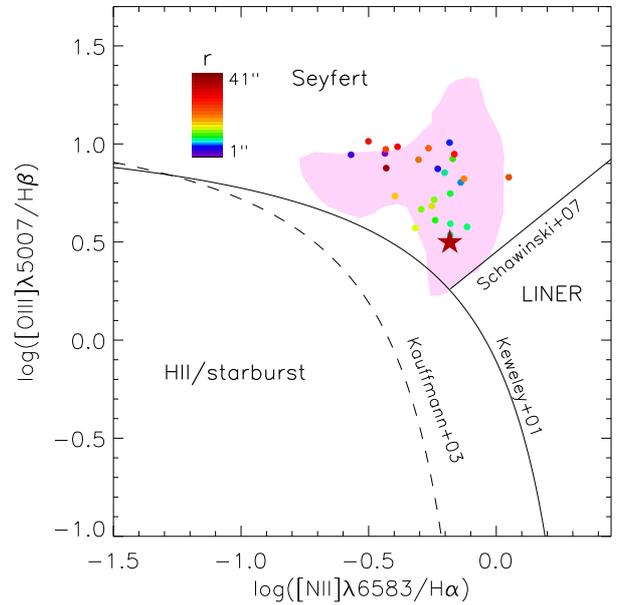}
}
\caption{BPT diagram for the spectra along $PA=144\degr$ and $226\degr$, different colours correspond to radial distances along the slit according to the scale box.
The red star marks emission-line  fitting results for the spectrum shown in Fig.~\ref{fig_spec} integrated over $r=-82..-72$ arcsec. The division lines between star-forming galaxies, Seyfert and LINER nucleus are taken from \citet{Kauffmann2003}, \citet{Kewley2001}, and \citet{Schawinski2007}. The violet filled regions mark the envelope of the main part of points in the $3\times5$ arcsec central field according to \citet{Freitas2018}.
}
\label{fig_BPT}
\end{figure}

In order to study a gas ionization origin, we made two cross-sections with SCORPIO-2 in the long-slit spectroscopy mode: in $PA=144\degr$ \citep[along the brightest part of the \OIII structure N to the nucleus near the position already studied spectroscopically in][]{Meaburn1989}  and along   $PA=226\degr$ (through the nucleus and  the brightest part of the distant filament). The example of the latter spectrum  is shown in Fig.~\ref{fig_spec}. We were able to detect the brightest emission lines (H$\alpha$, H$\beta$, \NII, and \OIII) in the distant emission knot at $r\approx77$ arcsec. Figure~\ref{fig_BPT} presents the emission-line ratios for the [N~{\sc ii}]$\lambda6583$/\Ha and [OIII~{\sc ii}]$\lambda6583$/\Hb diagram \citep*[the BPT diagram after][]{Bald1981} based on the Gaussian fitting of the emission lines along both slits with the 2.8-arcsec  sampling. Only the regions with $S/N>2$ in all the lines were included in this plot. All the measurements in the galactic disc region ($r<40$ arcsec)  including the above-mentioned emission knot are located in the Seyfert region  of the BPT diagram implying that even at these large distances, the AGN radiation is dominated in  gas excitation similar to the \citet{Fischer2017} data for the inner part of the EELR.

{ The ratio of the narrow-band emission-line  images  reveals  several compact emission knots in the galactic disc ($r=20-50$ arcsec) with domination of the Balmer line emission: \OIII/\Ha$\leq0.3$ that corresponds to \OIII/\Hb$\leq1$. However, SCORPIO-2 slits did not cross these HII regions.}

 \section{Discussion}

Our new data reveal a complex structure of the EELR in Mrk~6 consisting of a gaseous disc co-rotating with the galaxy's stellar disc and off-plane ionizing filaments extended  to a projected  distance of 40 kpc. The BPT diagram  suggests that the AGN hard radiation  is a dominant source of gas ionization. However, the corresponding line ratios lie near the separation borders of Syefert/LINER/Starburts, therefore, the ionization balance should be checked. Based on the  technique and assumptions described in details by \citet{keel2017} (`case B' recombination,  similarity of the densest gaseous clouds at various radii $r$), we can write the ratio of the required emission rate of ionizing photons ($Q_{ion}$) needed to produce the observed \Ha surface brightness $F$ to the same rate powered by the observed-epoch nuclear \Ha flux  $(F_{nuc})$ as: $a_{ion}=Q_{ion}/Q_{nuc}=4\pi r^2F/F_{nuc}$, where $r$ is in the same  units as a square for $F$ (px, arcsec, etc.). Applying  this relation to our narrow-band \Ha image with $F_{nuc}=5.9\cdot10^{-13} \mbox{erg}\,\mbox{s}^{-1}\,\mbox{cm}^{-2}$ estimated in the $6$-arcsec diameter aperture, we obtained $a_{ion}<$1--3 in all the radii. The real value of $a_{ion}$ might be even  smaller, because we significantly underestimate  $Q_{nuc}$ in this calculation neglecting the dust absorption and circumnuclear gaseous clouds covering factor. Therefore, we can conclude that the present AGN radiation is enough for ionization of the whole EELR. 

%We ruled out the formation of extended filaments  
Formation of the extended filaments via the nuclear activity (AGN outflow, star burst wind, or jet-cloud interaction) contradicts its observed properties: (i) the gas is dynamically cold according to the FPI data on the velocity dispersion in the filaments  which is  similar  to those in the unperturbed regions in the disc ($\sigma<30-50\kms$); (ii) the radio jet does not extend to the corresponding distances; (iii) we do not see a  counter-structure expected for a symmetric outflow; (iv) the spatial scale  of a possible outflow is typical of a more powerful active nucleus (radio--loud QSO).

An alternative supposition is an external matter orbiting  orthogonally to the stellar disc according to our analysis of the \Ha velocity field. In this case, the origin of the NE filaments in Mrk~6  is similar to that in gaseous clouds in tidal tails and debris caught in the ionization cone of the AGN, e.g., the sample compiled by \citet{Keel2012}. The kinematics of these structures (the domination of rotation, a small velocity dispersion) according to \citet{Keel2015} is in agreement with the picture observed in Mrk~6, as well as with spatial scales. For instance, the EELR in NGC 5972 is detected at $r\approx50$ kpc together with the stellar tail \citep{Keel2012}.

\begin{figure}
\centerline{
\includegraphics[width=0.45\textwidth ]{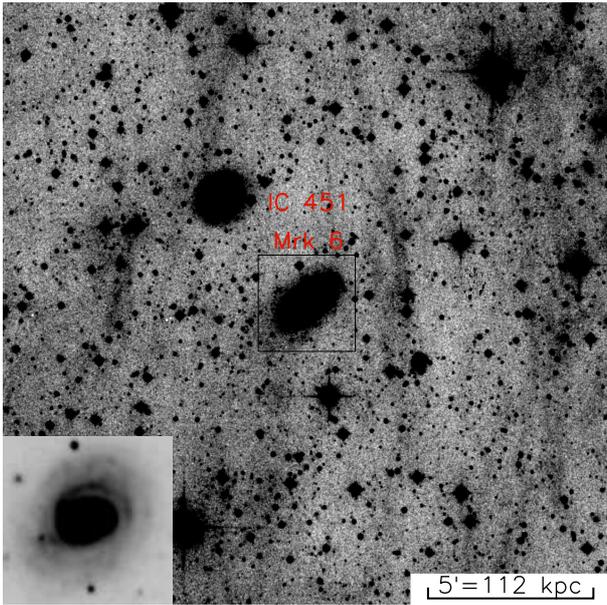}
}
\caption{Deep image of the Mrk~6  environment in the {\it g}-band obtained with the 1-m Schmidt telescope of BAO. 
%the most faint structures correspond to the surface brightness $28^m\mbox{arcsec}^2$. 
The square marks the field shown in Fig.~\ref{fig_maps}. The inset shows the zoomed $1\farcm5$ field centered at IC~451.}
\label{fig_BAO}
\end{figure}

With the aim to find possible stellar counterparts of the gaseous structure, we performed deep imaging observation of the Mrk~6 environment  at the 1-m Scmidt telescope of BAO (Section~\ref{sec_obs}) having a very low level of the scattered light as compared to the 6-m telescope  data essential for detecting faint structures. The reached  $1\sigma$ noise level corresponds to  surface brightnesses in the $g,r,i$ filters: 28.1, 27.5, and 26.9 mag~arcsec$^{-2}$, respectively. The deepest $g$-band image shown in Fig.~\ref{fig_BAO} reveals a large-scale filamentary system of the Milky Way cirruses elongated  mostly in the N-S direction with a typical brightness of $\sim26.5$ mag~arcsec$^{-2}$ seen in all the filters. However, we do not see any structures  which might be interpreted as tidal stellar streams related to the accreted gas. Only one possible source of accretion has appeared in our deep images -- the IC~451 galaxy which is similar in luminosity to Mrk~6 at a projected distance of $~98$~kpc. Moreover, the galaxy's disc has an asymmetric one-armed structure (the inset in Fig.~\ref{fig_BAO}) that could be related to a tidal interaction in the past; the Mrk 6 gaseous filaments are located exactly along the line between these galaxies. Can we consider the kinematically decoupled gaseous system in Mrk~6 as the first stage of formation of a gaseous polar ring in agreement with a popular accretion scenario  \citep{ReshetnikovSotnikova1997,BournaudCombes2003}? We have only one strong argument against this hypothesis. The  IC~451 systemic velocity equal to $5123\pm53\kms$ (according to NED) is $~500\kms$ smaller than that in Mrk~6; it is too high for a gravitationally bound  pair of galaxies. Even if we observe the first passage  of IC~451 near  Mrk~6, it is difficult to understand  how the accreted gas dramatically changed its own angular momentum to reach  observed line-of-sight velocities of $\sim6000\kms$  that by $700\kms$ differs from those of IC~451. 

{ The observed difference of the line-of-sight velocities in the  system Mrk~6--filaments gas--IC451 also  rules out the hypothesis that the elongation of the most extended filaments in the NE direction is caused by a recent collision between the galaxies. Indeed, one of the best illustrations of such interaction appears in the Virgo cluster: the gas stripping from the spiral galaxy NGC~4438 after its collision with the M~86 elliptical galaxy. The result is a system of   \Ha\ filaments  connecting both companions of the pair \citep{Kenney2008}. However, in this case a smooth gradient of the line-of-sight gas velocities is expected, i.e., the value of filament velocities  should be between systemic velocities of both companions, as it is observed in M86/NGC~4438 according to \citet{Kenney2008}. While this predicted picture  significantly contradicts the velocity distribution in the Mrk~6 environment. Moreover, as we already mentioned in Sec.~\ref{sec_morph}, the  disc of Mrk~6 has a  very regular  symmetric  structure that is inconsistent with a recent flyby interaction with a massive galaxy \citep[see][as an example of simulations]{Mapelli2008}.

Another way of formation of a one-sided `tail' of warm emitted gas is a ram-pressure stripping caused by the hot intergalactic medium in clusters or even in groups. The Seyfert galaxy NGC~4388 is a well-known example of such filaments \citep[and references therein]{Yoshida2004}. However, we have no evidence of a  hot X ray halo around/near Mrk~6 in the literature. The published \textit{Chandra} and \textit{XMM-Newton} data  reveal the hot gas only in the central bubbles  ($r<8$ kpc) related to the AGN-driven outflow \citep{Mingo2011}. }

{ Therefore,  all the available observed data are consistent with a supposition that} anisotropic radiation of the Seyfert nucleus collimated in a broad cone allows us to see a part of an off-plane gaseous structure orbiting around  Mrk~6. We restored its kinematics and the ionization state, while a possible source of gas accretion is ambiguous. We hope that new deep HI radio observations of this field will make it possible to reveal a clear picture of the gaseous environment and accretion processes around this well-known AGN.

\section{Acknowledgements}
The study  was supported by the Russian Science Foundation, project no. 17-12-01335. {We wish to thank an anonymous reviewer for constructive comments which have helped us to improve the paper.} This research has made use of the NASA/IPAC Extragalactic Database (NED) which is operated by the Jet Propulsion Laboratory, California Institute of Technology, under contract with the National Aeronautics and Space Administration. 
%The Pan-STARRS1 Surveys (PS1) and the PS1 public science archive have been made possible through contributions by the Institute for Astronomy, the University of Hawaii, the Pan-STARRS Project Office, the Max-Planck Society and its participating institutes, the Max Planck Institute for Astronomy, Heidelberg and the Max Planck Institute for Extraterrestrial Physics, Garching, The Johns Hopkins University, Durham University, the University of Edinburgh, the Queen's University Belfast, the Harvard-Smithsonian Center for Astrophysics, the Las Cumbres Observatory Global Telescope Network Incorporated, the National Central University of Taiwan, the Space Telescope Science Institute, the National Aeronautics and Space Administration under Grant No. NNX08AR22G issued through the Planetary Science Division of the NASA Science Mission Directorate, the National Science Foundation Grant No. AST-1238877, the University of Maryland, Eotvos Lorand University (ELTE), the Los Alamos National Laboratory, and the Gordon and Betty Moore Foundation.
\bibliographystyle{mnras}
\bibliography{galaxies}

\begin{thebibliography}{}
\makeatletter
\relax
\def\mn@urlcharsother{\let\do\@makeother \do\$\do\&\do\#\do\^\do\_\do\%\do\~}
\def\mn@doi{\begingroup\mn@urlcharsother \@ifnextchar [ {\mn@doi@}
  {\mn@doi@[]}}
\def\mn@doi@[#1]#2{\def\@tempa{#1}\ifx\@tempa\@empty \href
  {http://dx.doi.org/#2} {doi:#2}\else \href {http://dx.doi.org/#2} {#1}\fi
  \endgroup}
\def\mn@eprint#1#2{\mn@eprint@#1:#2::\@nil}
\def\mn@eprint@arXiv#1{\href {http://arxiv.org/abs/#1} {{\tt arXiv:#1}}}
\def\mn@eprint@dblp#1{\href {http://dblp.uni-trier.de/rec/bibtex/#1.xml}
  {dblp:#1}}
\def\mn@eprint@#1:#2:#3:#4\@nil{\def\@tempa {#1}\def\@tempb {#2}\def\@tempc
  {#3}\ifx \@tempc \@empty \let \@tempc \@tempb \let \@tempb \@tempa \fi \ifx
  \@tempb \@empty \def\@tempb {arXiv}\fi \@ifundefined
  {mn@eprint@\@tempb}{\@tempb:\@tempc}{\expandafter \expandafter \csname
  mn@eprint@\@tempb\endcsname \expandafter{\@tempc}}}

\bibitem[\protect\citeauthoryear{{Afanasiev} \& {Moiseev}}{{Afanasiev} \&
  {Moiseev}}{2005}]{AfanasievMoiseev2005}
{Afanasiev} V.~L.,  {Moiseev} A.~V.,  2005, \mn@doi [Astronomy Letters]
  {10.1134/1.1883351}, \href
  {http://adsabs.harvard.edu/abs/2005AstL...31..194A} {31, 194}

\bibitem[\protect\citeauthoryear{{Afanasiev} \& {Moiseev}}{{Afanasiev} \&
  {Moiseev}}{2011}]{SCORPIO2}
{Afanasiev} V.~L.,  {Moiseev} A.~V.,  2011, Baltic Astronomy, \href
  {http://adsabs.harvard.edu/abs/2011BaltA..20..363A} {20, 363}

\bibitem[\protect\citeauthoryear{{Afanasiev} \& {Silchenko}}{{Afanasiev} \&
  {Silchenko}}{1991}]{AfanaSil1991}
{Afanasiev} V.~L.,  {Silchenko} O.~K.,  1991, Astrofizicheskie Issledovaniia,
  \href {http://adsabs.harvard.edu/abs/1991AISAO..33..132A} {33, 132}

\bibitem[\protect\citeauthoryear{{Afanasiev}, {Dodonov}  \&
  {Moiseev}}{{Afanasiev} et~al.}{2001}]{Afanasiev2001}
{Afanasiev} V.~L.,  {Dodonov} S.~N.,   {Moiseev} A.~V.,  2001, in {Ossipkov}
  L.~P.,  {Nikiforov} I.~I.,  eds, Stellar Dynamics: from Classic to Modern.
  p.~103

\bibitem[\protect\citeauthoryear{{Afanasiev}, {Popovi{\'c}}, {Shapovalova},
  {Borisov}  \& {Ili{\'c}}}{{Afanasiev} et~al.}{2014}]{Afanasiev2014}
{Afanasiev} V.~L.,  {Popovi{\'c}} L.~{\v C}.,  {Shapovalova} A.~I.,  {Borisov}
  N.~V.,   {Ili{\'c}} D.,  2014, \mn@doi [\mnras] {10.1093/mnras/stu231}, \href
  {http://adsabs.harvard.edu/abs/2014MNRAS.440..519A} {440, 519}

\bibitem[\protect\citeauthoryear{{Baldwin}, {Phillips}  \&
  {Terlevich}}{{Baldwin} et~al.}{1981}]{Bald1981}
{Baldwin} J.~A.,  {Phillips} M.~M.,   {Terlevich} R.,  1981, \mn@doi [\pasp]
  {10.1086/130766}, \href {http://adsabs.harvard.edu/abs/1981PASP...93....5B}
  {93, 5}

\bibitem[\protect\citeauthoryear{{Bournaud} \& {Combes}}{{Bournaud} \&
  {Combes}}{2003}]{BournaudCombes2003}
{Bournaud} F.,  {Combes} F.,  2003, \mn@doi [\aap]
  {10.1051/0004-6361:20030150}, \href
  {http://adsabs.harvard.edu/abs/2003A%26A...401..817B} {401, 817}

\bibitem[\protect\citeauthoryear{{Cantalupo}, {Arrigoni-Battaia}, {Prochaska},
  {Hennawi}  \& {Madau}}{{Cantalupo} et~al.}{2014}]{Cantalupo2014}
{Cantalupo} S.,  {Arrigoni-Battaia} F.,  {Prochaska} J.~X.,  {Hennawi} J.~F.,
  {Madau} P.,  2014, \mn@doi [\nat] {10.1038/nature12898}, \href
  {http://adsabs.harvard.edu/abs/2014Natur.506...63C} {506, 63}

\bibitem[\protect\citeauthoryear{{Capetti}, {Axon}, {Kukula}, {Macchetto},
  {Pedlar}, {Sparks}  \& {Boksenberg}}{{Capetti} et~al.}{1995}]{Capetti1995}
{Capetti} A.,  {Axon} D.~J.,  {Kukula} M.,  {Macchetto} F.,  {Pedlar} A.,
  {Sparks} W.~B.,   {Boksenberg} A.,  1995, \mn@doi [\apjl] {10.1086/309784},
  \href {http://adsabs.harvard.edu/abs/1995ApJ...454L..85C} {454, L85}

\bibitem[\protect\citeauthoryear{{Dodonov}, {Kotov}, {Movsesyan}  \&
  {Gevorkyan}}{{Dodonov} et~al.}{2017}]{Dodonov2017}
{Dodonov} S.~N.,  {Kotov} S.~S.,  {Movsesyan} T.~A.,   {Gevorkyan} M.,  2017,
  \mn@doi [Astrophysical Bulletin] {10.1134/S1990341317040113}, \href
  {http://adsabs.harvard.edu/abs/2017AstBu..72..473D} {72, 473}

\bibitem[\protect\citeauthoryear{{Doroshenko}, {Sergeev}, {Klimanov}, {Pronik}
  \& {Efimov}}{{Doroshenko} et~al.}{2012}]{Doroshenko2012}
{Doroshenko} V.~T.,  {Sergeev} S.~G.,  {Klimanov} S.~A.,  {Pronik} V.~I.,
  {Efimov} Y.~S.,  2012, \mn@doi [\mnras] {10.1111/j.1365-2966.2012.20843.x},
  \href {http://adsabs.harvard.edu/abs/2012MNRAS.426..416D} {426, 416}

\bibitem[\protect\citeauthoryear{{Egorov}, {Lozinskaya}, {Moiseev}  \&
  {Smirnov-Pinchukov}}{{Egorov} et~al.}{2018}]{Egorov2018}
{Egorov} O.~V.,  {Lozinskaya} T.~A.,  {Moiseev} A.~V.,   {Smirnov-Pinchukov}
  G.~V.,  2018, \mn@doi [\mnras] {10.1093/mnras/sty1158}, \href
  {http://adsabs.harvard.edu/abs/2018MNRAS.478.3386E} {478, 3386}

\bibitem[\protect\citeauthoryear{{Eracleous} \& {Halpern}}{{Eracleous} \&
  {Halpern}}{1993}]{Eracleous}
{Eracleous} M.,  {Halpern} J.~P.,  1993, \mn@doi [\apj] {10.1086/172689}, \href
  {http://adsabs.harvard.edu/abs/1993ApJ...409..584E} {409, 584}

\bibitem[\protect\citeauthoryear{{Fischer} et~al.,}{{Fischer}
  et~al.}{2017}]{Fischer2017}
{Fischer} T.~C.,  et~al., 2017, \mn@doi [\apj] {10.3847/1538-4357/834/1/30},
  \href {http://adsabs.harvard.edu/abs/2017ApJ...834...30F} {834, 30}

\bibitem[\protect\citeauthoryear{{Freitas} et~al.,}{{Freitas}
  et~al.}{2018}]{Freitas2018}
{Freitas} I.~C.,  et~al., 2018, \mn@doi [\mnras] {10.1093/mnras/sty303}, \href
  {http://adsabs.harvard.edu/abs/2018MNRAS.476.2760F} {476, 2760}

\bibitem[\protect\citeauthoryear{{Kauffmann} et~al.,}{{Kauffmann}
  et~al.}{2003}]{Kauffmann2003}
{Kauffmann} G.,  et~al., 2003, \mn@doi [\mnras]
  {10.1111/j.1365-2966.2003.07154.x}, \href
  {http://adsabs.harvard.edu/abs/2003MNRAS.346.1055K} {346, 1055}

\bibitem[\protect\citeauthoryear{{Keel} et~al.,}{{Keel}
  et~al.}{2012}]{Keel2012}
{Keel} W.~C.,  et~al., 2012, \mn@doi [\mnras]
  {10.1111/j.1365-2966.2011.20101.x}, \href
  {http://adsabs.harvard.edu/abs/2012MNRAS.420..878K} {420, 878}

\bibitem[\protect\citeauthoryear{{Keel} et~al.,}{{Keel}
  et~al.}{2015}]{Keel2015}
{Keel} W.~C.,  et~al., 2015, \mn@doi [\aj] {10.1088/0004-6256/149/5/155}, \href
  {http://adsabs.harvard.edu/abs/2015AJ....149..155K} {149, 155}

\bibitem[\protect\citeauthoryear{{Keel} et~al.,}{{Keel}
  et~al.}{2017}]{keel2017}
{Keel} W.~C.,  et~al., 2017, \mn@doi [\apj] {10.3847/1538-4357/835/2/256},
  \href {http://adsabs.harvard.edu/abs/2017ApJ...835..256K} {835, 256}

\bibitem[\protect\citeauthoryear{{Kenney}, {Tal}, {Crowl}, {Feldmeier}  \&
  {Jacoby}}{{Kenney} et~al.}{2008}]{Kenney2008}
{Kenney} J.~D.~P.,  {Tal} T.,  {Crowl} H.~H.,  {Feldmeier} J.,   {Jacoby}
  G.~H.,  2008, \mn@doi [\apjl] {10.1086/593300}, \href
  {http://adsabs.harvard.edu/abs/2008ApJ...687L..69K} {687, L69}

\bibitem[\protect\citeauthoryear{{Kewley}, {Dopita}, {Sutherland}, {Heisler}
  \& {Trevena}}{{Kewley} et~al.}{2001}]{Kewley2001}
{Kewley} L.~J.,  {Dopita} M.~A.,  {Sutherland} R.~S.,  {Heisler} C.~A.,
  {Trevena} J.,  2001, \mn@doi [\apj] {10.1086/321545}, \href
  {http://adsabs.harvard.edu/abs/2001ApJ...556..121K} {556, 121}

\bibitem[\protect\citeauthoryear{{Khachikian} \& {Weedman}}{{Khachikian} \&
  {Weedman}}{1971}]{Khachikian}
{Khachikian} E.~E.,  {Weedman} D.~W.,  1971, Astrofizika, \href
  {http://adsabs.harvard.edu/abs/1971Afz.....7..389K} {7, 389}

\bibitem[\protect\citeauthoryear{{Kharb}, {O'Dea}, {Baum}, {Colbert}  \&
  {Xu}}{{Kharb} et~al.}{2006}]{Kharb2006}
{Kharb} P.,  {O'Dea} C.~P.,  {Baum} S.~A.,  {Colbert} E.~J.~M.,   {Xu} C.,
  2006, \mn@doi [\apj] {10.1086/507945}, \href
  {http://adsabs.harvard.edu/abs/2006ApJ...652..177K} {652, 177}

\bibitem[\protect\citeauthoryear{{Kharb}, {O'Dea}, {Baum}, {Hardcastle},
  {Dicken}, {Croston}, {Mingo}  \& {Noel-Storr}}{{Kharb}
  et~al.}{2014}]{Kharb2014}
{Kharb} P.,  {O'Dea} C.~P.,  {Baum} S.~A.,  {Hardcastle} M.~J.,  {Dicken} D.,
  {Croston} J.~H.,  {Mingo} B.,   {Noel-Storr} J.,  2014, \mn@doi [\mnras]
  {10.1093/mnras/stu421}, \href
  {http://adsabs.harvard.edu/abs/2014MNRAS.440.2976K} {440, 2976}

\bibitem[\protect\citeauthoryear{{Kukula}, {Holloway}, {Pedlar}, {Meaburn},
  {Lopez}, {Axon}, {Schilizzi}  \& {Baum}}{{Kukula} et~al.}{1996}]{Kukula1996}
{Kukula} M.~J.,  {Holloway} A.~J.,  {Pedlar} A.,  {Meaburn} J.,  {Lopez} J.~A.,
   {Axon} D.~J.,  {Schilizzi} R.~T.,   {Baum} S.~A.,  1996, \mn@doi [\mnras]
  {10.1093/mnras/280.4.1283}, \href
  {http://adsabs.harvard.edu/abs/1996MNRAS.280.1283K} {280, 1283}

\bibitem[\protect\citeauthoryear{{Lintott} et~al.,}{{Lintott}
  et~al.}{2009}]{Lintott2009}
{Lintott} C.~J.,  et~al., 2009, \mn@doi [\mnras]
  {10.1111/j.1365-2966.2009.15299.x}, \href
  {http://adsabs.harvard.edu/abs/2009MNRAS.399..129L} {399, 129}

\bibitem[\protect\citeauthoryear{{Magnier} et~al.,}{{Magnier}
  et~al.}{2016}]{Magnier2016}
{Magnier} E.~A.,  et~al., 2016, preprint, \href
  {http://adsabs.harvard.edu/abs/2016arXiv161205242M} {} (\mn@eprint {arXiv}
  {1612.05242})

\bibitem[\protect\citeauthoryear{{Mapelli}, {Moore}  \&
  {Bland-Hawthorn}}{{Mapelli} et~al.}{2008}]{Mapelli2008}
{Mapelli} M.,  {Moore} B.,   {Bland-Hawthorn} J.,  2008, \mn@doi [\mnras]
  {10.1111/j.1365-2966.2008.13421.x}, \href
  {http://adsabs.harvard.edu/abs/2008MNRAS.388..697M} {388, 697}

\bibitem[\protect\citeauthoryear{{Meaburn}, {Whitehead}  \& {Pedlar}}{{Meaburn}
  et~al.}{1989}]{Meaburn1989}
{Meaburn} J.,  {Whitehead} M.~J.,   {Pedlar} A.,  1989, \mn@doi [\mnras]
  {10.1093/mnras/241.1.1P}, \href
  {http://adsabs.harvard.edu/abs/1989MNRAS.241P...1M} {241, 1P}

\bibitem[\protect\citeauthoryear{{Mingo}, {Hardcastle}, {Croston}, {Evans},
  {Hota}, {Kharb}  \& {Kraft}}{{Mingo} et~al.}{2011}]{Mingo2011}
{Mingo} B.,  {Hardcastle} M.~J.,  {Croston} J.~H.,  {Evans} D.~A.,  {Hota} A.,
  {Kharb} P.,   {Kraft} R.~P.,  2011, \mn@doi [\apj]
  {10.1088/0004-637X/731/1/21}, \href
  {http://adsabs.harvard.edu/abs/2011ApJ...731...21M} {731, 21}

\bibitem[\protect\citeauthoryear{{Moiseev}}{{Moiseev}}{2014}]{Moiseev2014}
{Moiseev} A.~V.,  2014, \mn@doi [Astrophysical Bulletin]
  {10.1134/S1990341314010015}, \href
  {http://adsabs.harvard.edu/abs/2014AstBu..69....1M} {69, 1}

\bibitem[\protect\citeauthoryear{{Morganti}}{{Morganti}}{2017}]{Morganti2017}
{Morganti} R.,  2017, \mn@doi [Nature Astronomy] {10.1038/s41550-017-0223-0},
  \href {http://adsabs.harvard.edu/abs/2017NatAs...1..596M} {1, 596}

\bibitem[\protect\citeauthoryear{{Osterbrock} \& {Koski}}{{Osterbrock} \&
  {Koski}}{1976}]{Osterbrock}
{Osterbrock} D.~E.,  {Koski} A.~T.,  1976, \mn@doi [\mnras]
  {10.1093/mnras/176.1.61P}, \href
  {http://adsabs.harvard.edu/abs/1976MNRAS.176P..61O} {176, 61P}

\bibitem[\protect\citeauthoryear{{Reshetnikov} \& {Sotnikova}}{{Reshetnikov} \&
  {Sotnikova}}{1997}]{ReshetnikovSotnikova1997}
{Reshetnikov} V.,  {Sotnikova} N.,  1997, \aap, \href
  {http://adsabs.harvard.edu/abs/1997A%26A...325..933R} {325, 933}

\bibitem[\protect\citeauthoryear{{Schawinski}, {Thomas}, {Sarzi}, {Maraston},
  {Kaviraj}, {Joo}, {Yi}  \& {Silk}}{{Schawinski}
  et~al.}{2007}]{Schawinski2007}
{Schawinski} K.,  {Thomas} D.,  {Sarzi} M.,  {Maraston} C.,  {Kaviraj} S.,
  {Joo} S.-J.,  {Yi} S.~K.,   {Silk} J.,  2007, \mn@doi [\mnras]
  {10.1111/j.1365-2966.2007.12487.x}, \href
  {http://adsabs.harvard.edu/abs/2007MNRAS.382.1415S} {382, 1415}

\bibitem[\protect\citeauthoryear{{Smirnova} \& {Moiseev}}{{Smirnova} \&
  {Moiseev}}{2010}]{Smirnova2010}
{Smirnova} A.,  {Moiseev} A.,  2010, \mn@doi [\mnras]
  {10.1111/j.1365-2966.2009.15635.x}, \href
  {http://adsabs.harvard.edu/abs/2010MNRAS.401..307S} {401, 307}

\bibitem[\protect\citeauthoryear{{Yoshida} et~al.,}{{Yoshida}
  et~al.}{2004}]{Yoshida2004}
{Yoshida} M.,  et~al., 2004, \mn@doi [\aj] {10.1086/380221}, \href
  {http://adsabs.harvard.edu/abs/2004AJ....127...90Y} {127, 90}

\makeatother
\end{thebibliography}

\label{lastpage}

\end{document}